\documentclass[11pt]{article}
\usepackage{amsmath,amssymb,amsfonts,bbm}
\usepackage{amsmath,amssymb,amsfonts,bbm}
\usepackage{multirow}
\usepackage[table,xcdraw]{xcolor}
\usepackage{graphicx}
\usepackage{caption}
\usepackage[font={scriptsize}]{subcaption}
\usepackage[export]{adjustbox}
\usepackage{setspace}
\usepackage{float}
\usepackage{soul,color}
\sethlcolor{yellow}
\newcommand{\RNum}[1]{\uppercase\expandafter{\romannumeral #1\relax}}

\usepackage{hyperref}
\hypersetup{
   colorlinks   = true, 
    urlcolor     = blue, 
    linkcolor    =  blue, 
    citecolor   =  blue
}
\newcommand{\be}{\begin{equation}}
\newcommand{\ee}{\end{equation}}
\newcommand{\bea}{\begin{eqnarray}}
\newcommand{\eea}{\end{eqnarray}}
\newcommand{\ba}{\begin{array}}
\newcommand{\ea}{\end{array}}

\newcommand{\htwo}{h_{2,1}}
\newcommand{\M}{\mathcal{M}}
\newcommand{\N}{\mathcal{N}}

\newcommand{\K}{\mathcal{K}}

\long\def\symbolfootnote[#1]#2{\begingroup%
\def\thefootnote{\fnsymbol{footnote}}\footnote[#1]{#2}\endgroup}

\begin{document}

\thispagestyle{empty}\vspace{40pt}

\hfill{}

\vspace{128pt}

\begin{center}
    \textbf{\Large{Interpreting the Cosmic History of the Universe Through Five-Dimensional Supergravity}}\\
    \vspace{40pt}

    Moataz H. Emam$^{a,\, b}$\symbolfootnote[1]{\tt moataz.emam@cortland.edu} and Safinaz Salem$^{c}$\symbolfootnote[2]{\tt safinaz.salem@azhar.edu.eg}
\end{center}

    \vspace{3pt}
    \begin{changemargin}{0.5in}{0in}
    \begin{flushleft}
       \begin{spacing}{1.0}
    $^a$ \textit{\small  Department of Physics, SUNY College at Cortland, Cortland, New York 13045, USA,}\\
    $^b$ \textit{\small  University of Science and technology, Zewail City of Science and Technology, Giza 12578, Egypt,}\\
    $^c$ \textit{\small  Department of Physics, Faculty of Science, Al Azhar University, Cairo 11765, Egypt.}\\
        \end{spacing}
    \end{flushleft}
    \end{changemargin}

\vspace{6pt}

\begin{abstract}
    Through modeling the universe as a symplectic 3-brane embedded in the bulk of $\N=2$ five-dimensional ungauged supergravity theory, the entire evolution of the universe can be interpreted from inflation to late-time acceleration without introducing an inflaton nor a cosmological constant. The time dependence of the brane is strongly correlated to the complex structure moduli of the underlying Calabi-Yau submanifold and the bulk effects. The solutions to the field equations are found by exploiting the theory's symplectic structure where the time evolution is similar to our universe according to the latest data of the Planck mission. Our results present a new explanation for the nature of dark energy mainly based on the topology of the subspace and the existence of a fifth extra dimension.
\end{abstract}

\newpage



\vspace{15pt}

\pagebreak

\section{Introduction}
Cosmological evolution incorporates many mysteries. For instance, we assume with high confidence that our observable universe came
to exist in an event known as the Big Bang some $13.842$ billion years ago \cite{Bennett:2013}, but little is known about its initial conditions. In other words what caused the Big Bang to happen? Did the universe, in
a sense, have any choice about coming to exist? The general theory of relativity (GR) implies that it
started with a singularity, but did it? It is also known that shortly after the Big Bang, a period
of rapid acceleration; known as ‘inflation,’ caused the universe to expand very rapidly in a very
short period \cite{Tsujikawa}. Following a period of the so-called ‘graceful exit,’ decelerated expansion occurred. This last is the only epoch whose reasons are fully understood: namely the gravitational
interaction of the universe on itself following the standard Friedmann equations. Then, roughly
around our current time, a period of slow accelerated expansion started, as discovered by the Supernova Cosmology Project \cite{SupernovaCosmologyProject:1998vns}, and the Supernova Search Team \cite{SupernovaSearchTeam:1998fmf}.
The `dark energy' refers to the mysterious source of negative pressure which drives the universe for that expansion. In Einstein's gravitational theory this energy is accounted for by adding a term to the field equation called the cosmological constant 
($\Lambda$) \cite{Caldwell:2009}. But unfortunately, the observational constraint on the density of the dark energy is
$|\rho_\Lambda| < 10^{-27} ~\text{kg/m}^3 \sim 10^{-47} ~\text{GeV}^4$ \cite{Carroll:1992, Rugh:2000ji}, while the cosmological constant in GR receives many corrections from the quantum vacuum energy whose 
energy density ($\rho_{th}$) theoretically exceeds the observational
bound by at least 120 orders of magnitude. That means
$\Lambda$ is of order $ \sim 9.9834 \times 10^{-120}~ \Lambda_{th}$.

On the other hand, in recent years much attention has been drawn to five-dimensional $\mathcal{D}=5$ $\N=2$ supergravity theories for many reasons. For instance, they are so relevant to the AdS/CFT correspondence \cite{Nishimura:2000wj}, and they have a vital role in M-theory compactification on the Calabi-Yau manifold with cosmological applications \cite{Jarv:2004}. 
There are many proposed models of `brane-cosmology' in which 
the universe is considered as a 3-brane (supersymmetric or non-supersymmetric) embedded in a higher dimensional spacetime \cite{Maartens:2010}. Some of these models explain inflation based on the existence of a scalar field in the early universe 
called the `inflaton' \cite{Linde:1994yf}. Other models focus on the late-time acceleration of the universe \cite{Koyama:2008}.
However, it seems that studies of the whole cosmic evolution of the universe by a single model are rare.

In this paper, we aim to explain the entire cosmic history of the universe during its different acceleration phases from inflation 
till the present accelerated expansion.  
We model the universe as a symplectic 3-brane embedded in the bulk of $\N=2$ $\mathcal{D}=5$ supergravity that has been produced from the dimensional reduction of 
$\mathcal{D}=11$ $\N=1$ supergravity over Calabi-Yau 3-fold \cite{Cadavid:1995bk}. The brane is filled by dust matter and radiation, whilst its cosmological constant vanishes, and only the extra fifth dimension 
has a cosmological constant called ($\tilde{\Lambda}$). 
When we numerically solve the modified field equations, we find that the flow norm of the moduli
of the complex structure of the underlying Calabi-Yau manifold starts with very large values and
decays rapidly (without any imposed initial conditions) \cite{Dine:2006, Bodeker:2006}. We also find that there is a strong
correlation between the moduli and the dynamics of the brane, especially since it is correlated to an
early epoch of rapid expansion, i.e., inflation. Whereas the late-time acceleration occurs because
of the presence of the bulk's cosmological constant mediated by the moduli flow norm. 
The solutions are BPS (Bogomol'nyi-Prasad-Sommerfield) \cite{Stelle:2008} in the sense they partially preserve supersymmetry, and they are related to the initial conditions of the brane-universe.
To fit the recent observations of the $\Lambda$CDM model \cite{Ostriker:1995}, 
the bulk should be a de Sitter space with a tiny positive cosmological constant $ (0.02 <\tilde{\Lambda} <0.03) ~ [\text{Gyr}^{-2}]$.
In other words, an expanding universe does not need a cosmological constant on the brane to explain its late-time acceleration nor the so-called inflaton to describe the early-time inflation. The whole cosmic evolution of the universe can be explained by topological effects and the existence of an extra dimension.
Also, we solve the modified Friedmann equations analytically when the bulk has a fixed size, which imposes a constraint on the 
density and the pressure in the brane and the cosmological constant in the bulk.

It is worth mentioning that in a previous work of one of us, a similar model has been explored
but the brane was vacuous \cite{Emam:2015laa}, so it was considered a toy model of our universe. In a further study
\cite{Emam:2020} we investigated a brane filled with dust and radiation separately and only the inflation era of
the universe has been considered.

The paper is organized as follows: In section (\RNum{2}) we review the $\mathcal{D}=5$ supergravity theory formulated in the symplectic structure. In section (\RNum{3}) we introduce our metric for the model, solve the modified Friedmann equations numerically, and show the Hubble parameter and the scale factor of the brane coincide with that of the $\Lambda$CDM model over the scanned region of the model's free parameter.  
In section (\RNum{4}), we declare how the entire cosmic evolution of the brane-universe agrees with our universe's expansion history
through the different epochs, from the early inflation to the late-time acceleration.
In section (\RNum{5}), we solve the field equations analytically.

\section{Five dimensional $\N=2$ supergravity} \label{theory}

The ungauged five dimensional $\N=2$ supergravity theory contains two sets of matter fields; the vector multiplets, which we set to zero, and our main interest: the \emph{hypermultiplets}. These are composed of the \emph{universal hypermultiplet} $\left(\phi, \sigma, \zeta^0, \tilde \zeta_0\right)$; where $\phi$ is the universal axion, and the dilaton $\sigma$ is proportional to the volume of the underlying Calabi-Yau manifold $\M$. The remaining hypermultiplet scalars are $\left(z^i, z^{\bar i}, \zeta^i, \tilde \zeta_i: i=1,\ldots, \htwo\right)$, where the $z$'s are the complex structure moduli of $\M$, and $\htwo$ is the Hodge number determining the dimensions of the manifold $\M_C$ of the Calabi-Yau's complex structure moduli\footnote{A `bar' over an index denotes complex conjugation}. The fields $\left(\zeta^I, \tilde\zeta_I: I=0,\ldots,\htwo\right)$ are the axions, which define a symplectic vector space (see \cite{Emam:2010kt} for a review and more references). The axions are defined as components of the symplectic vector
\be\label{DefOfSympVect}
   \left| \Xi  \right\rangle  = \left( {\begin{array}{*{20}c}
   {\,\,\,\,\,\zeta ^I }  \\
   -{\tilde \zeta _I }  \\
    \end{array}} \right),
\ee
such that the symplectic scalar product is defined by, for example,
$
    \left\langle {{\Xi }}
 \mathrel{\left | {\vphantom {{\Xi } \Xi }}
 \right. \kern-\nulldelimiterspace}
 {\Xi } \right\rangle   = \zeta^I \tilde \zeta_I  - \tilde \zeta_I
 \zeta^I.\label{DefOfSympScalarProduct}
$
A transformation in symplectic space can be defined by
\be
 \left\langle {d\Xi } \right|\mathop {\bf\Lambda} \limits_ \wedge  \left| {\star d\Xi } \right\rangle
  = 2\left\langle {{d\Xi }}
 \mathrel{\left | {\vphantom {{d\Xi } V}}
 \right. \kern-\nulldelimiterspace}
 {V} \right\rangle \mathop {}\limits_ \wedge  \left\langle {{\bar V}}
 \mathrel{\left | {\vphantom {{\bar V} {\star d\Xi }}}
 \right. \kern-\nulldelimiterspace}
 {{\star d\Xi }} \right\rangle  + 2G_{i\bar j} \left\langle {{d\Xi }}
 \mathrel{\left | {\vphantom {{d\Xi } {U_{\bar j} }}}
 \right. \kern-\nulldelimiterspace}
 {{U_{\bar j} }} \right\rangle \mathop {}\limits_ \wedge  \left\langle {{U_i }}
 \mathrel{\left | {\vphantom {{U_i } {\star d\Xi }}}
 \right. \kern-\nulldelimiterspace}
 {{\star d\Xi }} \right\rangle  - i\left\langle {d\Xi } \right.\mathop |\limits_ \wedge  \left. {\star d\Xi } \right\rangle,\label{DefOfRotInSympSpace}
\ee
where $d$ is the spacetime exterior derivative, $\star$ is the five dimensional Hodge duality operator, and $G_{i\bar j}$ is a special K\"{a}hler metric on $\M_C$. The symplectic basis vectors $\left| V \right\rangle $, $\left| {U_i } \right\rangle $ and their complex conjugates are defined by
\be
    \left| V \right\rangle  = e^{\frac{\K}{2}} \left( {\begin{array}{*{20}c}
   {Z^I }  \\
   {F_I }  \\
    \end{array}} \right),\,\,\,\,\,\,\,\,\,\,\,\,\,\,\,\left| {\bar V} \right\rangle  = e^{\frac{\K}{2}} \left( {\begin{array}{*{20}c}
   {\bar Z^I }  \\
   {\bar F_I }  \\
    \end{array}} \right)\label{DefOfVAndVBar}
\ee

\noindent where $\K$ is the K\"{a}hler potential on $\M_C$, $\left( {Z,F} \right)$ are the periods of the Calabi-Yau's holomorphic volume form, and

\bea
    \left| {U_i } \right\rangle  &=& \left| \nabla _i V
    \right\rangle=\left|\left[ {\partial _i  + \frac{1}{2}\left( {\partial _i \K} \right)} \right] V \right\rangle \nonumber\\
    \left| {U_{\bar i} } \right\rangle  &=& \left|\nabla _{\bar i}  {\bar V} \right\rangle=\left|\left[ {\partial _{\bar i}  + \frac{1}{2}\left( {\partial _{\bar i} \K} \right)} \right] {\bar V}
    \right\rangle\label{DefOfUAndUBar}
\eea
where the derivatives are with respect to the moduli $\left(z^i, z^{\bar i}\right)$. In this language, the bosonic part of the action is given by:
\bea
    S_5  &=& \int\limits_5 {\left[ {\mathcal{R}\star \mathbf{1} - \frac{1}{2}d\sigma \wedge\star d\sigma  - G_{i\bar j} dz^i \wedge\star dz^{\bar j} } \right.}  + e^\sigma   \left\langle {d\Xi } \right|\mathop {\bf\Lambda} \limits_ \wedge  \left| {\star d\Xi } \right\rangle\nonumber\\
    & &\left. {\quad\quad\quad\quad\quad\quad\quad\quad\quad\quad\quad\quad\quad - \frac{1}{2} e^{2\sigma } \left[ {d\phi + \left\langle {\Xi } \mathrel{\left | {\vphantom {\Xi  {d\Xi }}} \right. \kern-\nulldelimiterspace} {{d\Xi }}    \right\rangle} \right] \wedge \star\left[ {d\phi + \left\langle {\Xi } \mathrel{\left | {\vphantom {\Xi  {d\Xi }}} \right. \kern-\nulldelimiterspace} {{d\Xi }}    \right\rangle} \right] } \right].\label{action}
\eea

Where $\mathcal{R}$ is the curvature scalar of the five-dimensional metric $g_{MN}$, $M,N=0,...,4$.  The usual $\delta S = 0$ gives the following field equations for the hypermultiplets scalar fields:
\bea
    \left( {\Delta \sigma } \right)\star \mathbf{1} + e^\sigma   \left\langle {d\Xi } \right|\mathop {\bf\Lambda} \limits_ \wedge  \left| {\star d\Xi } \right\rangle -   e^{2\sigma }\left[ {d\phi + \left\langle {\Xi } \mathrel{\left | {\vphantom {\Xi  {d\Xi }}} \right. \kern-\nulldelimiterspace} {{d\Xi }}    \right\rangle} \right]\wedge\star\left[ {d\phi + \left\langle {\Xi } \mathrel{\left | {\vphantom {\Xi  {d\Xi }}} \right. \kern-\nulldelimiterspace} {{d\Xi }}    \right\rangle} \right] &=& 0\label{DilatonEOM}\\
    \left( {\Delta z^i } \right)\star \mathbf{1} + \Gamma _{jk}^i dz^j  \wedge \star dz^k  + \frac{1}{2}e^\sigma  G_{i\bar j}  {\partial _{\bar j} \left\langle {d\Xi } \right|\mathop {\bf\Lambda} \limits_ \wedge  \left| {\star d\Xi } \right\rangle} &=& 0 \nonumber\\
    \left( {\Delta z^{\bar i} } \right)\star \mathbf{1} + \Gamma _{\bar j\bar k}^{\bar i} dz^{\bar j}  \wedge \star dz^{\bar k}  + \frac{1}{2}e^\sigma  G_{\bar ij}  {\partial _j \left\langle {d\Xi } \right|\mathop {\bf\Lambda} \limits_ \wedge  \left| {\star d\Xi } \right\rangle}  &=& 0\label{ZZBarEOM} \\
    d^{\dag} \left\{ {e^\sigma  \left| {{\bf\Lambda} d\Xi } \right\rangle  - e^{2\sigma } \left[ {d\phi + \left\langle {\Xi }
    \mathrel{\left | {\vphantom {\Xi  {d\Xi }}}\right. \kern-\nulldelimiterspace} {{d\Xi }} \right\rangle } \right]\left| \Xi  \right\rangle } \right\} &=& 0\label{AxionsEOM}\\
    d^{\dag} \left[ {e^{2\sigma } d\phi + e^{2\sigma } \left\langle {\Xi } \mathrel{\left | {\vphantom {\Xi  {d\Xi }}} \right. \kern-\nulldelimiterspace} {{d\Xi }}    \right\rangle} \right] &=&    0\label{aEOM}
\eea
where $d^\dagger$ is the $\mathcal{D}=5$ adjoint exterior derivative, $\Delta$ is the Laplace-de Rahm operator and $\Gamma _{jk}^i$ is a connection on $\M_C$. The full action is symmetric under the following SUSY transformations:
\bea
 \delta _\epsilon  \psi ^1  &=& D \epsilon _1  + \frac{1}{4}\left\{ {i {e^{\sigma } \left[ {d\phi + \left\langle {\Xi }
 \mathrel{\left | {\vphantom {\Xi  {d\Xi }}}
 \right. \kern-\nulldelimiterspace} {{d\Xi }} \right\rangle } \right]}- Y} \right\}\epsilon _1  - e^{\frac{\sigma }{2}} \left\langle {{\bar V}}
 \mathrel{\left | {\vphantom {{\bar V} {d\Xi }}} \right. \kern-\nulldelimiterspace} {{d\Xi }} \right\rangle\epsilon _2  \nonumber\\
 \delta _\epsilon  \psi ^2  &=& D \epsilon _2  - \frac{1}{4}\left\{ {i {e^{\sigma } \left[ {d\phi + \left\langle {\Xi }
 \mathrel{\left | {\vphantom {\Xi  {d\Xi }}} \right. \kern-\nulldelimiterspace}
 {{d\Xi }} \right\rangle } \right]}- Y} \right\}\epsilon _2  + e^{\frac{\sigma }{2}} \left\langle {V}
 \mathrel{\left | {\vphantom {V {d\Xi }}} \right. \kern-\nulldelimiterspace} {{d\Xi }} \right\rangle \epsilon _1,  \label{SUSYGraviton}
\eea
\bea
  \delta _\epsilon  \xi _1^0  &=& e^{\frac{\sigma }{2}} \left\langle {V}
    \mathrel{\left | {\vphantom {V {\partial _\mu  \Xi }}} \right. \kern-\nulldelimiterspace} {{\partial _\mu  \Xi }} \right\rangle  \Gamma ^\mu  \epsilon _1  - \left\{ {\frac{1}{2}\left( {\partial _\mu  \sigma } \right) - \frac{i}{2} e^{\sigma } \left[ {\left(\partial _\mu \phi\right) + \left\langle {\Xi }
    \mathrel{\left | {\vphantom {\Xi  {\partial _\mu \Xi }}} \right. \kern-\nulldelimiterspace}
    {{\partial _\mu \Xi }} \right\rangle } \right]} \right\}\Gamma ^\mu  \epsilon _2  \nonumber\\
     \delta _\epsilon  \xi _2^0  &=& e^{\frac{\sigma }{2}} \left\langle {{\bar V}}
    \mathrel{\left | {\vphantom {{\bar V} {\partial _\mu  \Xi }}} \right. \kern-\nulldelimiterspace} {{\partial _\mu  \Xi }} \right\rangle \Gamma ^\mu  \epsilon _2  + \left\{ {\frac{1}{2}\left( {\partial _\mu  \sigma } \right) + \frac{i}{2} e^{\sigma } \left[ {\left(\partial _\mu \phi\right) + \left\langle {\Xi }
    \mathrel{\left | {\vphantom {\Xi  {\partial _\mu \Xi }}} \right. \kern-\nulldelimiterspace}
    {{\partial _\mu \Xi }} \right\rangle } \right]} \right\}\Gamma ^\mu  \epsilon
     _1,
\label{SUSYHyperon1}
\eea
and
\bea
     \delta _\epsilon  \xi _1^{\hat i}  &=& e^{\frac{\sigma }{2}} e^{\hat ij} \left\langle {{U_j }}
    \mathrel{\left | {\vphantom {{U_j } {\partial _\mu  \Xi }}} \right. \kern-\nulldelimiterspace} {{\partial _\mu  \Xi }} \right\rangle \Gamma ^\mu  \epsilon _1  - e_{\,\,\,\bar j}^{\hat i} \left( {\partial _\mu  z^{\bar j} } \right)\Gamma ^\mu  \epsilon _2  \nonumber\\
     \delta _\epsilon  \xi _2^{\hat i}  &=& e^{\frac{\sigma }{2}} e^{\hat i\bar j} \left\langle {{U_{\bar j} }}
    \mathrel{\left | {\vphantom {{U_{\bar j} } {\partial _\mu  \Xi }}} \right. \kern-\nulldelimiterspace} {{\partial _\mu  \Xi }} \right\rangle \Gamma ^\mu  \epsilon _2  + e_{\,\,\,j}^{\hat i} \left( {\partial _\mu  z^j } \right)\Gamma ^\mu  \epsilon    _1,
\label{SUSYHyperon2}
\eea
where $\left(\psi ^1, \psi ^2\right)$ are the two gravitini and $\left(\xi _1^I, \xi _2^I\right)$ are the hyperini. The quantity $Y$ is defined by:
\begin{equation}
    Y   = \frac{{\bar Z^I N_{IJ}  {d  Z^J }  -
    Z^I N_{IJ}  {d  \bar Z^J } }}{{\bar Z^I N_{IJ} Z^J
    }},\label{DefOfY}
\end{equation}
where $N_{IJ}  = \mathfrak{Im} \left({\partial_IF_J } \right)$. The $e$'s are the beins of the special K\"{a}hler metric $G_{i\bar j}$, the $\epsilon$'s are the five-dimensional $\N=2$ SUSY spinors and the $\Gamma$'s are the usual Dirac matrices. The covariant derivative $D$ is defined by the usual $D=dx^\mu\left( \partial _\mu   + \frac{1}{4}\omega _\mu^{\,\,\,\,\hat \mu\hat \nu} \Gamma _{\hat \mu\hat \nu}\right)\label{DefOfCovDerivative}$, where the $\omega$'s are the spin connections and the hatted indices are frame indices in a flat tangent space. Finally, the bulk's stress tensor is:
\bea
T_{\mu \nu }  &=& -\frac{1}{2}\left( {\partial _\mu  \sigma } \right)\left( {\partial _\nu  \sigma } \right) + \frac{1}{4}g_{\mu \nu } \left( {\partial _\alpha  \sigma } \right)\left( {\partial ^\alpha  \sigma } \right)
 + e^\sigma  \left\langle {\partial _\mu \Xi } \right|{\bf\Lambda} \left| {\partial _\nu \Xi } \right\rangle - \frac{1}{2}e^{\sigma } g_{\mu \nu }   \left\langle {\partial _\alpha \Xi } \right|{\bf\Lambda} \left| {\partial ^\alpha \Xi } \right\rangle \nonumber\\
  & &  - \frac{1}{2}e^{2\sigma } \left[ {\left( {\partial _\mu  \phi} \right) + \left\langle {\Xi }
 \mathrel{\left | {\vphantom {\Xi  {\partial _\mu  \Xi }}}
 \right. \kern-\nulldelimiterspace}
 {{\partial _\mu  \Xi }} \right\rangle } \right]\left[ {\left( {\partial _\nu  \phi} \right) + \left\langle {\Xi }
 \mathrel{\left | {\vphantom {\Xi  {\partial _\nu  \Xi }}}
 \right. \kern-\nulldelimiterspace}
 {{\partial _\nu  \Xi }} \right\rangle } \right]
  +  \frac{1}{4}e^{2\sigma } g_{\mu \nu } \left[ {\left( {\partial _\alpha  \phi} \right) + \left\langle {\Xi }
 \mathrel{\left | {\vphantom {\Xi  {\partial _\alpha  \Xi }}}
 \right. \kern-\nulldelimiterspace}
 {{\partial _\alpha  \Xi }} \right\rangle } \right]\left[ {\left( {\partial ^\alpha  \phi} \right) + \left\langle {\Xi }
 \mathrel{\left | {\vphantom {\Xi  {\partial ^\alpha  \Xi }}}
 \right. \kern-\nulldelimiterspace}
 {{\partial ^\alpha  \Xi }} \right\rangle } \right]\nonumber\\
 & & - G_{i\bar j} \left( {\partial _\mu  z^i } \right)\left( {\partial _\nu  z^{\bar j} } \right) + \frac{1}{2}g_{\mu \nu } G_{i\bar j} \left( {\partial _\alpha  z^i } \right)\left( {\partial ^\alpha  z^{\bar j} } \right).\label{StressTensor}
\eea

Where $\mu,\nu=0,...,3$. As our main interest is bosonic configurations that preserve \emph{some} supersymmetry, the stress tensor can be simplified by considering the vanishing of the supersymmetric variations (\ref{SUSYHyperon1},\ref{SUSYHyperon2}); satisfying the BPS condition on the brane. This gives
\be
    T_{\mu \nu }  = G_{i\bar j} \left( {\partial _\mu  z^i } \right)\left( {\partial _\nu  z^{\bar j} } \right) - \frac{1}{2}g_{\mu \nu } G_{i\bar j} \left( {\partial _\alpha  z^i } \right)\left( {\partial ^\alpha  z^{\bar j} } \right),\label{StressTens}
\ee
as was detailed out in \cite{Emam:2015laa}.

\section{Brane embedding and the field equations}

We construct a 3-brane that may be thought of as a flat Robertson-Walker universe embedded in five dimensions. This is mapped by the metric
\be\label{Metric}
    ds^2  = g_{MN}~ dx^M ~ dx^N =- dt^2  + a^2 \left( t \right) \left( {dr^2  + r^2 d\Omega ^2 } \right) + b^2 \left( t \right) dy^2,
\ee
where ${d\Omega ^2  = d\theta ^2  + \sin ^2 \left( \theta  \right)d\varphi ^2 }$, $a \left( t \right)$ is the usual Robertson-Walker scale factor, and $b \left( t \right)$ is a scale factor for the bulk dimension $y$. The brane is located at $y=0$ and we ignore all possible $y$-dependence of the warp factors as well as of the hypermultiplet bulk fields. We have explored this embedding in two earlier papers. In the first \cite{Emam:2015laa} an `empty' brane-world was analytically studied, and a correlation between the dynamics of the brane, \emph{i.e.} the time evolution of the scale factors, was found. This was further explored in \cite{Emam:2020} where two cases were considered: one with a dust-filled brane and the other with a radiation-filled brane. In both cases not only was the aforementioned correlation confirmed, leading to an expanding brane brought about by the complex structure moduli, but we have also found that the early decay of the norm of the moduli's flow velocity $G_{i\bar j} \dot z^i \dot z^{\bar j}$ directly leads to a period of rapid accelerated expansion of the brane. Not only does this imply an interesting underlying mechanism for future exploration, but it also implies that the complex structure moduli can play the role of the inflaton if one takes this embedding seriously from a cosmological perspective. We continue this possible application here by exploring the possible late-time effect that the moduli can have on the brane's dynamics by studying
a brane containing matter with density $\rho_m = \rho_{m0}/a^3$, and radiation plus neutrinos with density $\rho_r =  \rho_{r0}/a^4$. $\rho_{m0}$, and $\rho_{r0}$ are the current matter, and radiation plus neutrinos densities, respectively. Using the fact that for an ideal fluid, the typical radiation pressure is $p(t)= \rho(t)/3$, the following components of the brane's stress tensor are added to (\ref{StressTens}):
\bea
    T_{tt}^{{\rm Brane}}  &=& \rho_m\left(t\right) + \rho_r \left( t \right) = \frac{\rho_{m0}}{{a^3 }} + \frac{\rho_{r0}}{{a^4 }} \nonumber\\
    T_{rr}^{{\rm Brane}}  &=& a^2 p\left( t \right) = a^2 \frac{\rho_{r0}}{3a^4}\nonumber\\
    T_{\theta \theta }^{{\rm Brane}}  &=& a^2 r^2 p\left( t \right) = a^2 r^2 \frac{\rho_{r0}}{3a^4},\nonumber\\
    T_{\varphi \varphi }^{{\rm Brane}}  &=& a^2 r^2 \sin ^2 \theta p\left( t \right) = a^2 r^2 \sin ^2 \theta \frac{\rho_{r0}}{3a^4}.\label{Brane Stress}
\eea
Since the brane's stress tensor $T^{Brane}_{\mu\nu}$ does not arise from the theory's action Equ. (\ref{action}), the matter and radiation contents of the brane do not couple to the supersymmetry fermions, the gravitini, and the hyperini. So the supersymmetry variations Equ.  (\ref{SUSYGraviton}, \ref{SUSYHyperon1}, \ref{SUSYHyperon2}) are valid only in the bulk, while the brane's contents are also confined to the brane, since the fifth component of the brane's stress tensor $T^{Brane}_{yy} =0$.    
We consider the brane cosmological constant $\Lambda =0$ and show how the late-time acceleration can be produced only from the moduli  
and the bulk cosmological constant $\tilde \Lambda$ effects. 
Then Einstein's equations $G_{MN} + \Lambda g_{MN} = \kappa T_{MN} $, where $\kappa=\frac{8\pi G}{c^4}$ give the following Friedmann-like equations:
\bea
3 \left[ {\left( {\frac{{\dot a}}{a}} \right)^2  + \left( {\frac{{\dot a}}{a}} \right)\left( {\frac{{\dot b}}{b}} \right)} \right] &=& G_{i\bar j} \dot z^i \dot z^{\bar j}  + \rho \nonumber\\
 2\frac{{\ddot a}}{a} + \left( {\frac{{\dot a}}{a}} \right)^2  + \frac{{\ddot b}}{b} + 2\left( {\frac{{\dot a}}{a}} \right)\left( {\frac{{\dot b}}{b}} \right) &=&  - p - G_{i\bar j} \dot z^i \dot z^{\bar j}  \nonumber\\
3 \left[ {\frac{{\ddot a}}{a} + \left( {\frac{{\dot a}}{a}} \right)^2 } \right] &=&  \tilde\Lambda  - G_{i\bar j} \dot z^i \dot z^{\bar j} ,
\label{F-equ1}
\eea
where we consider $G=c=1$ units. Substitute by the current density parameters $ \Omega_0 \sim \rho_0/ 3 H^2$, the field equations become:
\bea
 \left[ {\left( {\frac{{\dot a}}{a}} \right)^2  + \left( {\frac{{\dot a}}{a}} \right)\left( {\frac{{\dot b}}{b}} \right)} \right] &=& G_{i\bar j} \dot z^i \dot z^{\bar j}  + H_0^2 \left ( \frac{{\Omega_{m0} }}{{a^3 }} +  \frac{\Omega_{r0} }{{a^4 }} \right) \nonumber\\
 2\frac{{\ddot a}}{a} + \left( {\frac{{\dot a}}{a}} \right)^2  + \frac{{\ddot b}}{b} + 2\left( {\frac{{\dot a}}{a}} \right)\left( {\frac{{\dot b}}{b}} \right) &=&  - H_0^2 \frac{\Omega_{r0}}{a^4} - G_{i\bar j} \dot z^i \dot z^{\bar j}  \nonumber\\
3 \left[ {\frac{{\ddot a}}{a} + \left( {\frac{{\dot a}}{a}} \right)^2 } \right] &=&  \tilde\Lambda  - G_{i\bar j} \dot z^i \dot z^{\bar j} ,
\eea
\begin{figure}[t]
  \begin{subfigure}[t]{.5\linewidth}
    \centering
    \includegraphics[width=1\columnwidth]{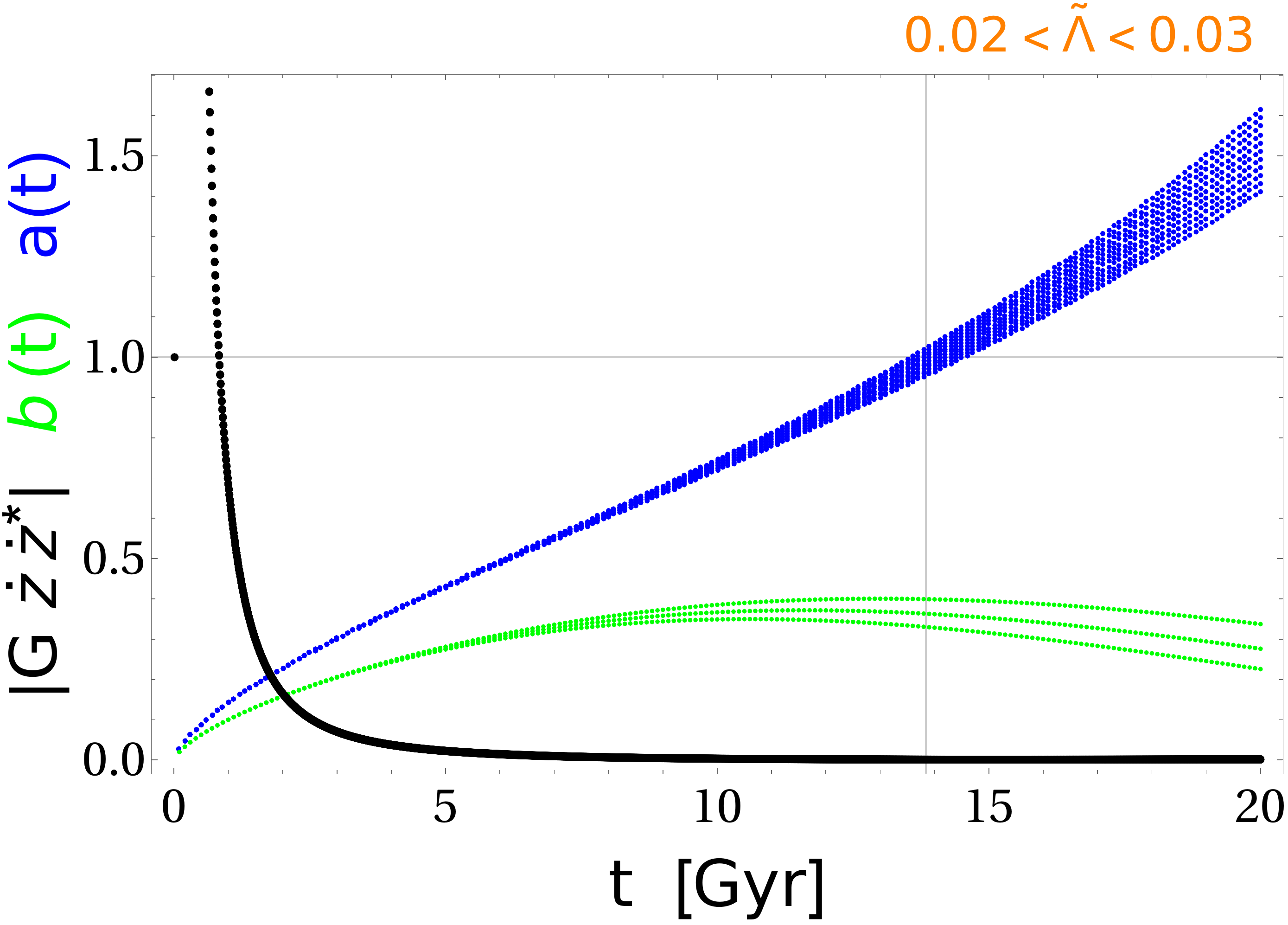}
    \caption{The brane's scale factor $a$ (blue), the bulk's scale factor $b$ (green), and  $\left| {G_{i\bar j} \dot z^i \dot z^{\bar j}} \right|$ (black) are plotted versus time for $0.02 < \tilde{\Lambda} < 0.03$. $a_0=1$, and $t_0= 13.842~ [\text{Gyr}]$.}
    \label{abz}
  \end{subfigure}
\qquad 
  \begin{subfigure}[t]{.5\linewidth}
    \centering
    \includegraphics[width=1\columnwidth]{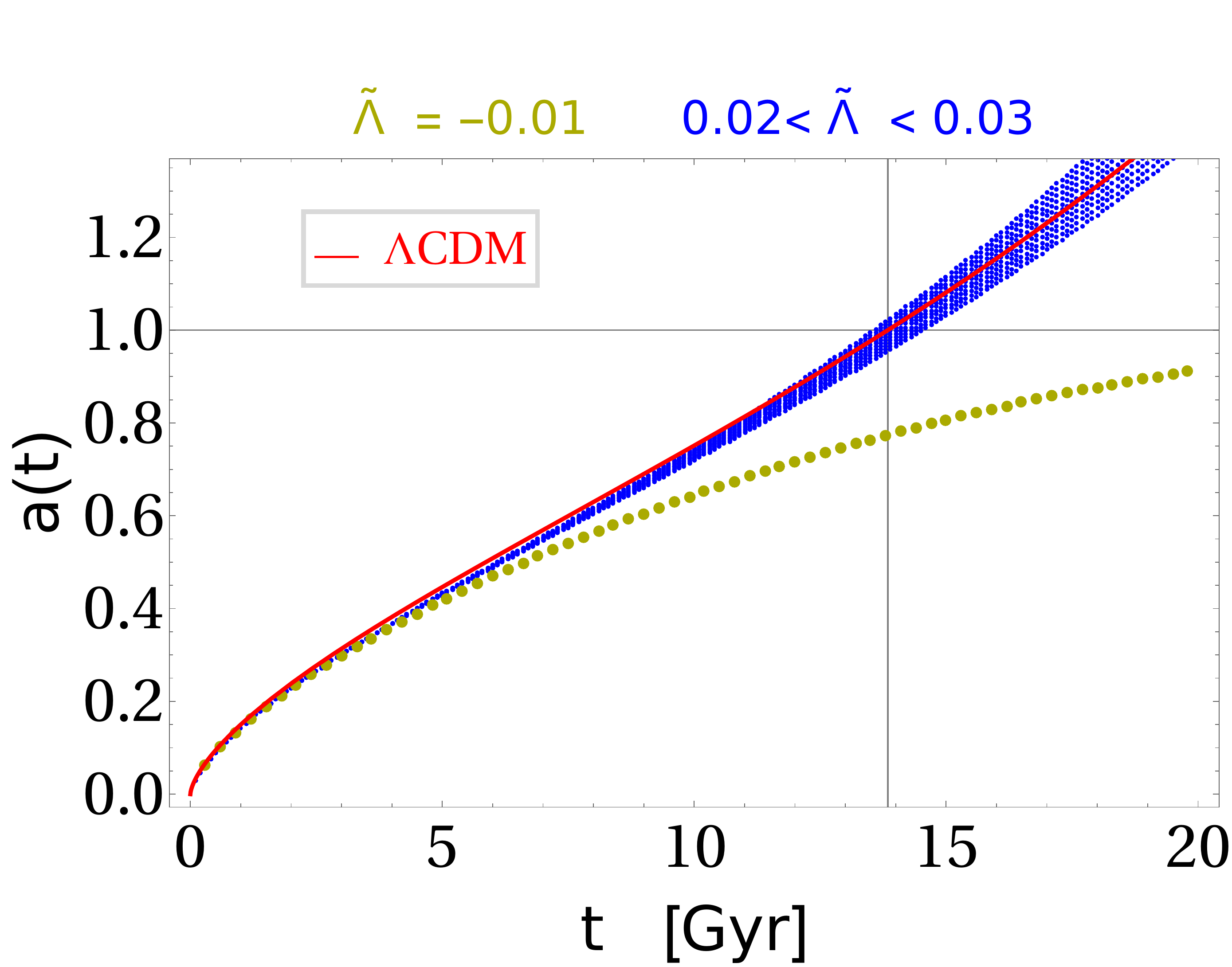}
    \caption{The brane's scale factor (blue) fits the scale factor of the $\Lambda$CDM model (red) for the range
    $0.02 < \tilde{\Lambda} < 0.03$. The yellow curve shows the brane's scale factor for $\tilde{\Lambda} = -0.01$.}
    \label{at}
  \end{subfigure}
\caption{}
\end{figure} 
The first and the second equations represent the Friedmann-like equations for the brane-universe, while the third equation represents the bulk. Solving those equations numerically to get the brane's and the bulk's scale factors $a$ and $b$, respectively, and the moduli's flow velocity $G_{i\bar j} \dot z^i \dot z^{\bar j}$ \cite{Salem:2022xdj}. We scan over a range of the bulk's cosmological constant $0.02 < \tilde{\Lambda} < 0.03$. Whilst the solutions are valid for a wide range of initial conditions fine-tuning. Here we take $a[0]=b[0] \sim 0.06$ and $a'[0]=b'[0] \sim 0.01$.
In our model, we consider the value of the dark energy density parameter $\Omega_{\Lambda}= 0$, and the current matter density parameter 
$\Omega_{m0} \sim 0.99$, so in our model the main components of the brane-universe are only matter and radiation. In the $\Lambda$CDM model according to the recent Planck mission data  based on CMB (The Cosmic Microwave Background) \cite{Planck:2018} $\Omega_{m0}= 0.3111$, the current radiation density parameter $\Omega_{r0}= 8.2 \times 10^{-5}$, and $\Omega_{\Lambda 0} = 0.6889$. The current value of the Hubble parameter is given by $H_0= 0.0686751 ~[\text{Gyr}^{-1}] \sim 2.176 \times 10^{-18}~ [\text{sec}^{-1}] $. The age of the universe is $t_0= 13.842~ [\text{Gyr}]$.
That amount of $\Omega_{m0}$ in the brane-universe means that the current dark matter density parameter 
$\Omega_{Dm0} \sim 0.94$ in the brane because the current baryonic
matter density parameter in our universe is $\Omega_{Bm0} = 0.0463 \pm 0.0024$. That huge amount of $\Omega_{Dm0}$ in the brane
may affect other cosmological observable like the baryonic acoustic oscillation or the galactic rotations curves.
However, next, we will show that the cosmic evolution of the brane-universe is similar to the 
expansion history of our universe according to BOSS collaboration data based on the baryon acoustic oscillation (BAO)
and the CMB constraints \cite{BOSS:2014hwf}.
\begin{figure}[t]
  \begin{subfigure}[t]{.5\linewidth}
    \centering
    \includegraphics[width=1\columnwidth]{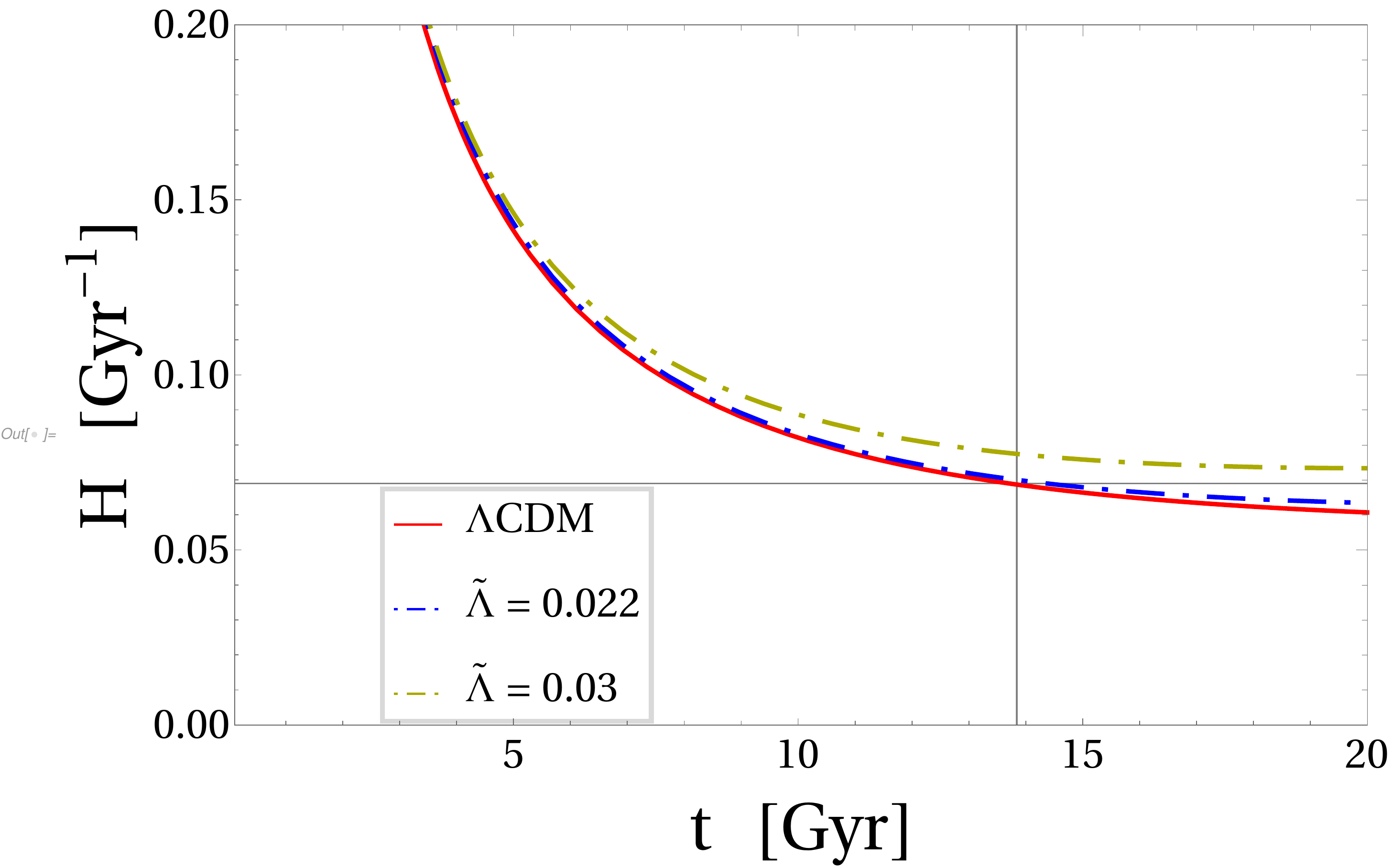}
    \caption{The Hubble parameter of $\Lambda$CDM (red), the brane's Hubble parameters for $\tilde{\Lambda}= 0.022 ~[\text{Gyr}^{-2}]$ (blue dashed) and $\tilde{\Lambda}= 0.03 ~[\text{Gyr}^{-2}]$ (yellow dashed) are plotted versus time. $H_0= 0.0689751 ~[\text{Gyr}^{-1}]$.}
    \label{Ht}
  \end{subfigure}
\qquad 
  \begin{subfigure}[t]{.5\linewidth}
    \centering
    \includegraphics[width=1\columnwidth]{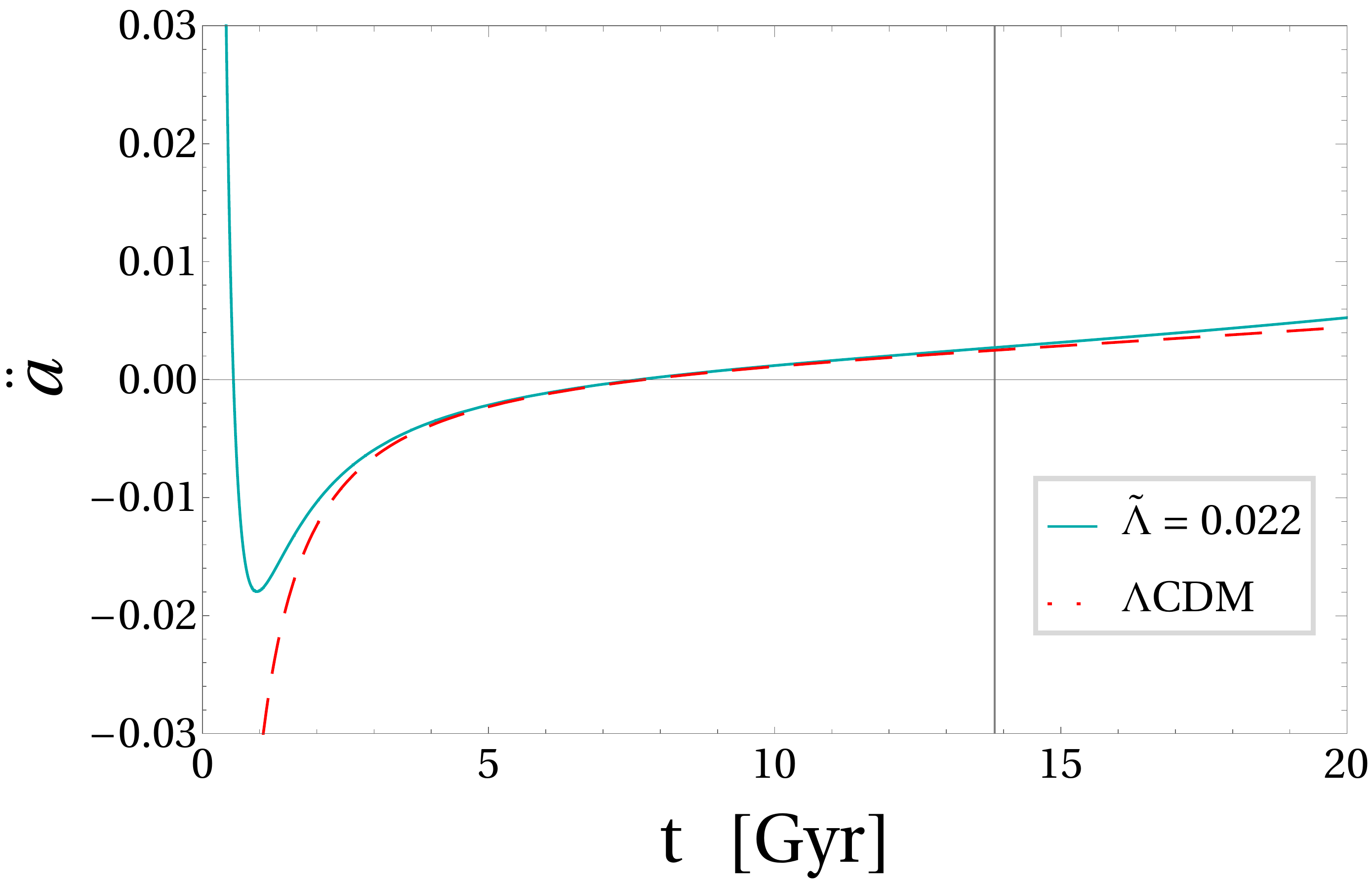}
     \caption{The acceleration of the brane's scale factor (blue) and the acceleration of the scale factor of $\Lambda$CDM (red dashed) are plotted versus time. Around $t=10 ~[\text{Gyr}]$ the accelerated expansion has been initiated.}
   \label{acel}
  \end{subfigure}
   \caption{}
\end{figure}
In Fig. (\ref{abz}) the brane's and the bulk's scale factors, and the absolute value of the moduli velocity norm are plotted versus time. Through these ranges $a[t_0] = 1 $ as the scale factor of the $\Lambda$CDM model. The bulk's scale factor varies by different values and it is inversely proportional to $\tilde{\Lambda}$. Also 
the correlation between $G_{i\bar j} \dot z^i \dot z^{\bar j}$ and $a(t)$ is clear, as the moduli are maximum $a(t)$ is minimum, and vise versa. Fig. (\ref{at}) shows that the scale factor of the brane (blue) fits and the $\Lambda$CDM's scale factor (red)
over the scanned parameter region, where the current value of the $\Lambda$CDM's scale factor $a(t_0)$ is normalized to $a_0=1$.
While for instance, an Anti-de Sitter bulk with a negative $\tilde{\Lambda}$ leads the brane-univesre for a deceleration.  
Fig. (\ref{Ht}) shows that brane's Hubble parameter (blue dashed) fits the Hubble parameter of the $\Lambda$CDM model (red) at $\tilde{\Lambda}= 0.022$. The horizontal line is at $H_0$. 
In Fig. (\ref{acel}) the acceleration of the brane's scale factor at $\tilde{\Lambda}=0.022$ (blue)
and the acceleration of the scale factor of $\Lambda$CDM (red dashed) are plotted versus time.
We can see the deceleration and the acceleration eras that happened after the big bang by $(3-5)$ Gyr 
and $(9-10)$ Gyr, respectively, according to the BOSS's data. Such that during the decelerated expansion  
the brane's acceleration was negative ($\ddot{a} < 0$), then around $10$ Gyr ($\ddot{a} > 0$) till the present time $(t_0)$.
Through those epochs the brane and the $\Lambda$CDM's accelerations coincide, while in the early times, the brane's acceleration
shows inflationary behavior ($\ddot{a} > 0$) which the $\Lambda$CDM model does not explain.
In the next graphs, we will show clearly the very early ages of the expansion.

\section{The cosmic history} 

\begin{figure}[t]
    \centering
    \includegraphics[width=0.6\columnwidth]{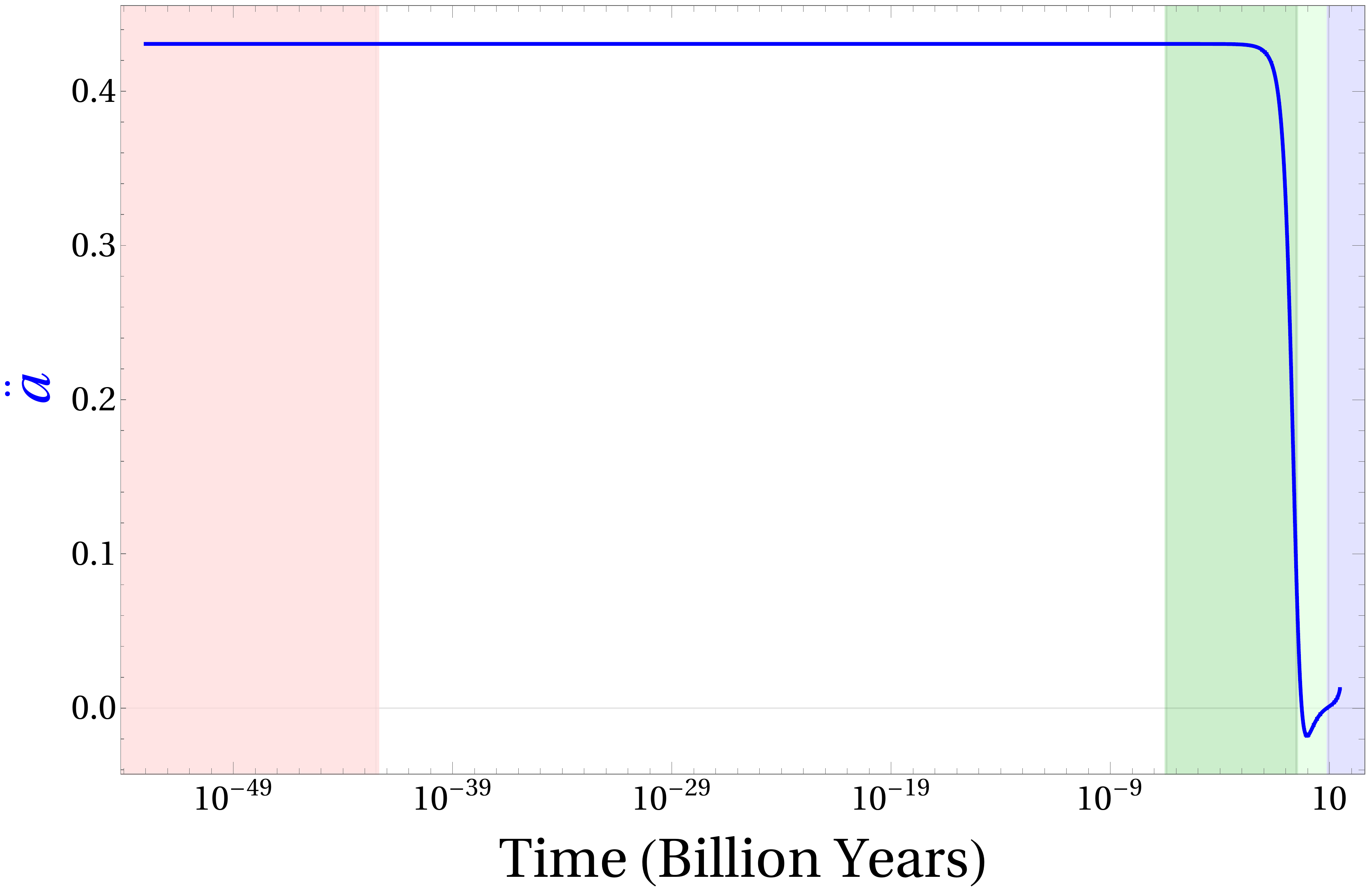}
    \caption{The acceleration of the brane's scale factor is plotted versus time on a logarithmic scale ($\tilde{\Lambda}=0.022$.). 
The pink area corresponds to the inflation era, the dark green area when the CMB happens, the light green area when the deceleration expansion happened, and the light blue area corresponds to the late-time acceleration era. }
    \label{acellog}
\end{figure}
\begin{figure}[t]
    \centering
    \includegraphics[width=0.6\columnwidth]{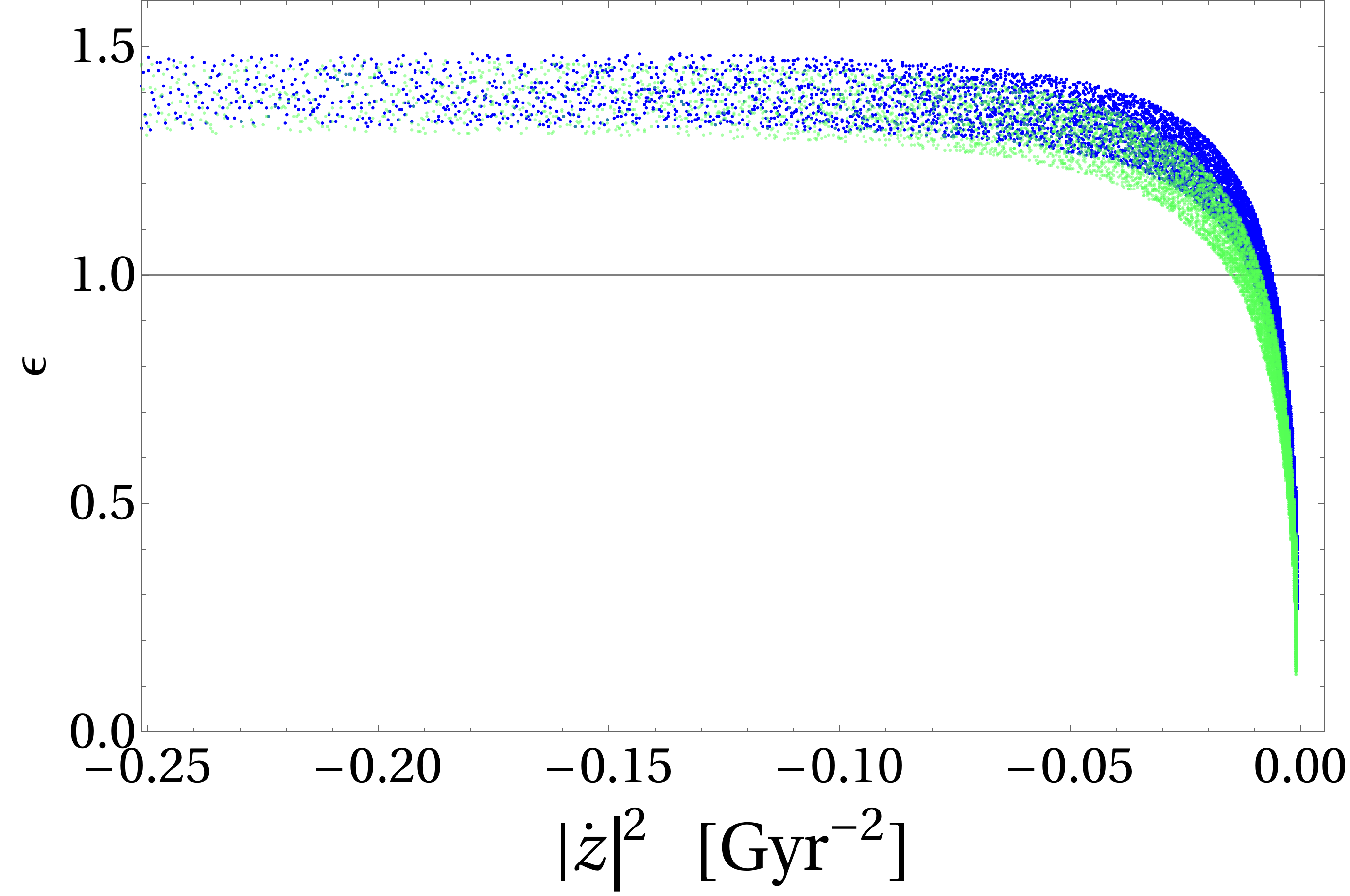}
    \caption{The slow-roll parameters are plotted against the moduli's flow velocity $ ({G_{i\bar j} \dot z^i \dot z^{\bar j}} )$ at 
$\tilde{\Lambda} =0.022$ (blue) and $\tilde{\Lambda} =0.03$ (green).}
\label{inf-pa}
\end{figure}

In Fig. (\ref{acellog}) the acceleration of the brane-universe is plotted versus time on a logarithmic scale. We can see that the early times are zoomed in, and the whole cosmic evolution of the universe is shown.  
The pink era is when the inflation started around $t \sim 10^{-53} ~[\text{Gyr}]$ and ended on $t \sim 10^{-43} ~[\text{Gyr}]$ where ($\ddot{a} > 0$).
Then the dark green era
when the CMB happened, the light green era when the decelerated expansion took place, and the light blue era 
corresponds to the late-time acceleration expansion. 
\begin{figure}[t]
    \centering
 \includegraphics[width=0.6\columnwidth]{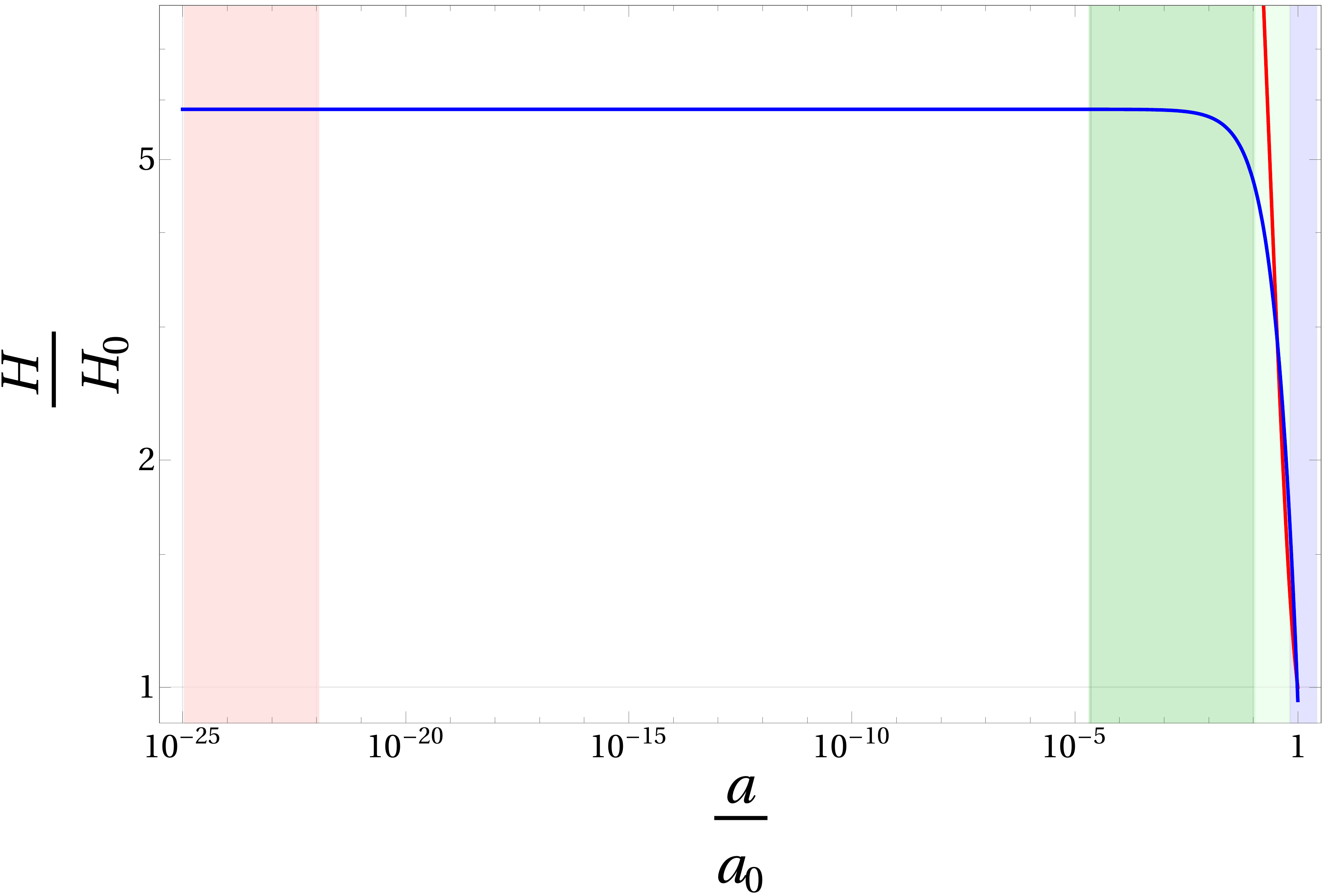}
    \caption{$ H(a)/H_0$ for the brane-universe (blue) and the $\Lambda$CDM model (red) are plotted versus 
the expansion scale factor. 
The pink area is a high regime era where $H>>H_0$, which corresponds to the inflation era, and the dark green area is when the CMB happens. 
Then $H/H_0$ starts to get close
to unity as the brane starts deceleration expansion through the light green area. The light blue area when the accelerated expansion 
started at $a\sim 0.7$ till the time beings. }
    \label{Ha}
\end{figure}
In Fig. (\ref{Ha}) the Hubble parameter $H(a)/H_0$ for the brane (blue) and the $\Lambda$CDM model (red)
are plotted as functions in the expansion scale factor $(a/a_0)$ on a logarithmic scale ($\tilde{\Lambda}=0.022$). The pink area corresponds to the inflation era, and
the dark green area starts when the CMB happened after the big bang by $t_{CMB}=380.000$ years and 
$a[t_{CMB}] \sim 2.3 \times 10^{-5}~[\text{Gyr}]$. The light green era corresponds to when the deceleration of the universe's expansion happened, and 
the light blue era corresponds to when the late-time acceleration happens after $(9-10)$ billion years of the big bang till time beings. That is according to the experimental data of our universe's expansion based on the baryon acoustic oscillation (BAO) reported by BOSS collaboration. 
We can see that $H(a)/H_0$ of the brane met that of the $\Lambda$CDM during the deceleration era. It is also shown that ($a/a_0$)
through the inflation period where there are enough number of 
e-folds $N=ln(a_e/a_0)$ (e stands for the end of the inflation) to solve the Horizon problem.  
Let us draw the attention that the values of $\tilde{\Lambda}$ are given in [$\text{Gyr}^{-2}$]. For the sake of comparison, the value of $\Lambda$ in $\Lambda$CDM is given by $\frac{8 \pi G}{c^2}~ \rho^{obs}_\Lambda~ \sim 9.7 \times 10^{-36} ~ \text{sec}^{-2} \sim 0.00969~ \text{Gyr}^{-2}$, while for instance $\tilde{\Lambda} = 0.022~ \text{Gyr}^{-2} = 5.5 \times 10^{-35}~ \text{sec}^{-2}$.
Fig. (\ref{inf-pa}) shows the relation of the moduli to the inflation slow-roll parameter $\epsilon = -\frac{\dot{H}}{H^2} $, where
$\epsilon$ is plotted against the moduli's flow velocity $|\dot{z}|^2 \equiv \left( {G_{i\bar j} \dot z^i \dot z^{\bar j}} \right)$ at $\tilde{\Lambda} =0.022$ (blue) and $\tilde{\Lambda} =0.03$ (green). We can see that when the moduli have larger values $\epsilon < 1$ at inflation which means that $H$ is varying slowly or there is an accelerated phase of expansion. Then when  $|\dot{z}|^2$ decrease the inflation ends, where $\epsilon > 1$.

\section{Finding analytic solution}
To find an analytic solution for the field equations (\ref{F-equ1}), we consider the bulk has a constant size, i.e., the bulk's scale factor is constant $(\dot{b}=0)$. The field equations simplify to:
\bea
3 \left( {\frac{{\dot a}}{a}} \right)^2 &=& G_{i\bar j} \dot z^i \dot z^{\bar j}  + \rho, \nonumber\\
 2\frac{{\ddot a}}{a} + \left( {\frac{{\dot a}}{a}} \right)^2  &=&  - p  - G_{i\bar j} \dot z^i \dot z^{\bar j},  \nonumber\\
 \left[ {\frac{{\ddot a}}{a} + \left( {\frac{{\dot a}}{a}} \right)^2 } \right] &=&  \tilde\Lambda  - G_{i\bar j} \dot z^i \dot z^{\bar j}.
\eea
Eliminating $(G_{i\bar j} \dot z^i \dot z^{\bar j})$  gives:
\bea
\nonumber 
p &=& \frac{2 \dot{a}^2}{a} + \frac{\ddot{a}}{a}+ \tilde{\Lambda} , \\
\rho &=&  \frac{4 \dot{a}^2}{a^2} + \frac{2\ddot{a}}{a}+ p.
\eea
Solving these couple of equations gives: 
\bea
\nonumber 
a(t) &=& c_2 \cos \left[ \sqrt{\frac{3}{2}} \sqrt{p-\rho} (t+ 2 c_1)  \right]^{1/3}, \\
&=&  c_2 \cos \left[ \sqrt{3} \sqrt{\tilde{\Lambda}-p} (t-  c_1)  \right]^{1/3}.
\eea
Where $c_1$ and $c_2$ are the integration constants. That sets a constrain on the density $\rho$ and the pressure $p$ in the brane, and $\tilde{\Lambda}$ in the bulk, $3 p -2 \rho=\tilde{\Lambda}.$ 

Another thing we want to declare about that class of models with a null cosmological constant ($\Lambda=0$) \cite{Jaime:2022}, that they can prompt the string theory to be a theory of quantum gravity. If our universe as described in this paper has no cosmological constant, the Swampland conjectures do not apply to the string theory as a low-energy theory of our universe.    

\section{Conclusion}
What is the origin of the late-time accelerated expansion of the universe? We find It
difficult to accept that its origin is the cosmological constant in Einstein's general
relativity because in GR the cosmological constant is related to the vacuum
energy density which is tremendously large compared to the observed energy
density (dark energy) that is responsible for the universe's accelerated expansion. In
this paper, we introduced a new explanation for dark energy by modeling the 
universe as a symplectic 3-brane embedded in a five-dimensional bulk of $\N=2$ supergravity. The
brane-universe has an early time inflation, followed by deceleration, then followed by a late-time acceleration 
adequate to our universe's cosmic evolution. Although the 3-brane has a zero
cosmological constant, solving the modified Friedmann's equations 
shows that the fields which drive the dynamics of the brane and the bulk are the complex
structure moduli of the Calabi-Yau manifold. The bulk should be a di-Sitter space
with a non-vanishing cosmological constant. The results correspond to the $\Lambda$CDM model and 
agree with the recent experimental data presented by BOSS collaboration about our universe's expansion history 
based on the baryon acoustic oscillation (BAO) combined with the CMB constraints. So our model gives an entire
explanation of the whole time evolution of the universe, from inflation to
the current time acceleration based on topological effects, and a fifth dimension. 
We would like to point out that in light of the results presented in this work, 
there are many other quests opened for future research, like
studying the dynamics of the Calabi-Yau manifold itself and its complex structure moduli,
investigating the moduli dependence on the extra fifth dimension, since here
only the time dependence is considered, fully analyzing the inflationary period,
what is the nature of the bulk's cosmological constant? Is the bulk's
energy density like the energy density proposed in the $\Lambda$CDM model?
Although the results agree with the BAO's data, what are the consequences of assuming a large amount of 
dark energy in the brane? Finally,
what are the implications of the analytic solutions of the field equations.


\begin{thebibliography}{999}

\bibitem{Bennett:2013}
C. L. Bennett, et al ``Nine-year Wilkinson Microwave anisotropy probe (WMAP) observations:
Final maps and results,'' The atrophysical journal supplement series, volume 208, number 2
(2013) doi:10.1088/0067-0049/208/2/20 [arXiv:1212.5225 [astro-ph]].

\bibitem{Tsujikawa}
S. Tsujikawa, ``Introductory review of cosmic inflation,'' hep-ph/0304257.

\bibitem{SupernovaCosmologyProject:1998vns}
S. Perlmutter, et al.,
``Measurements of $\Omega$ and $\Lambda$ from 42 high redshift supernovae,''
Astrophys. J.  {\bf 517} (1999) 565-586, arXiv:astro-ph/9812133.


\bibitem{SupernovaSearchTeam:1998fmf}
A.G. Riess, et al.,
``Observational evidence from supernovae for an accelerating universe and a cosmological constant,''
Astron. J.  { \bf 116} (1998) 1009-1038, arXiv:astro-ph/9805201.


\bibitem{Caldwell:2009}
R. R. Caldwell, and M. Kamionkowski, ``The physics of cosmic acceleration,'' Annual Review
of Nuclear and Particle Science, Vol. 59:397-429 (2009) doi:10.1146/annurev-nucl-010709-
151330.


\bibitem{Carroll:1992}
Sean M. Carroll, et al.,
``The Cosmological Constant,''
Ann. Rev. Astron. Astrophys. {\bf 30} (1992) 499-542.

\bibitem{Rugh:2000ji}
S.E. Rugh, H. Zinkernagel,
``The Quantum vacuum and the cosmological constant problem,''
Stud. Hist. Philos. Sci. B { \bf 33} (2002) 663-705, arXiv:hep-th/0012253.

\bibitem{Nishimura:2000wj}
M. Nishimura, 
``Conformal supergravity from the AdS/CFT correspondence,''
Nucl. Phys. B {\bf 588} (2000) 471–482, hep-th/0004179.


\bibitem{Jarv:2004}
L. Jarv, T. Mohaupt, and F. Saueressig,  
`` M-theory cosmologies from singular Calabi-Yau compactifications,''
JCAP {\bf 0402} (2004) 012, hep-th/0310174.



\bibitem{Maartens:2010}
R. Maartens and K. Koyama, ``Brane-World Gravity,'' Living Rev. Rel. 13, 5 (2010)
doi:10.12942/lrr-2010-5 [arXiv:1004.3962 [hep-th]].



\bibitem{Linde:1994yf}
A. D. Linde,
``Recent progress in inflationary cosmology,''
Lect. Notes Phys. {\bf 455} (1995) 363-372. Contribution to: International Workshop
on the Birth of the Universe and Fundamental Physics, 363-372, LLWI 1994, 72-109, hep-th/9410082. 

\bibitem{Koyama:2008}
K. Koyama, ``The cosmological constant and dark energy in braneworlds,''
Gen. Rel. Grav. {\bf 40}, 421 (2008), astro-ph/0706.1557.


\bibitem{Cadavid:1995bk}
A. C. Cadavid, A. Ceresole, R. D'Auria and S. Ferrara, 
``Eleven-dimensional supergravity compactified on Calabi-Yau threefolds,''
Phys. Lett. B {\bf 357} (1995) 76–80, hep-th/9506144.

\bibitem{Dine:2006}
M. Dine, R. Kitano, A. Morisse and Y. Shirman, ``Moduli decays and gravitinos,'' Phys. Rev.
D 73, 123518 (2006) [hep-ph/0604140].


\bibitem{Bodeker:2006}
D. Bodeker, ``Moduli decay in the hot early universe,'' JCAP 0606, 027 (2006) [hep-
ph/0605030].


\bibitem{Stelle:2008}
K. S. Stelle, `` BPS Branes in Supergravity,'' 
NATO Sci. Ser. C {\bf 530} (1999) 257 [hep-th/9803116].
D. N. Kabat and A. Rajaraman, ``Testing cosmological supersymmetry breaking,''
Phys. Lett. B {\bf 516}, 383 (2001) [hep-ph/0102309]; 
M. H. Emam, ``Zero-branes and the symplectic
hypermultiplets,'' Phys. Rev. D {\bf 86}, 045016 (2012) [hep-th/1208.3488].



\bibitem{Ostriker:1995}
J. P. Ostriker, Paul J. Steinhardt, 
``The Observational case for a low density universe with a
nonzero cosmological constant, '' Nature {\bf 377} (1995) 600.

\bibitem{Emam:2015laa}
M. H. Emam, ``BPS brane cosmology in N = 2 supergravity,'' Class. Quant. Grav. 32, no.
18, 185014 (2015) doi:10.1088/0264-9381/32/18/185014 [arXiv:1509.01651 [hep-th]].


\bibitem{Emam:2020}
M. Emam, H. H. Salah, and S. Salem,
``Brane-worlds and the Calabi- Yau complex structure moduli'',
Class. Quant. Grav. \textbf{37}, 195007 (2020)
\textit{hep-th/2005.10408}.



\bibitem{Emam:2010kt}
M. H. Emam, ``The Many symmetries of Calabi-Yau compactifications,'' Class. Quant. Grav.
27, 163001 (2010) [arXiv:1007.4847 [hep-th]].

\bibitem{Salem:2022xdj}
Safinaz Salem, Moataz H. Emam, and H.H. Salah,
`` The implications of $\N=2$ Supergravity Cosmology On the Topology of the Calabi-Yau Manifold,''
[arXiv:gr-qc/2204.13776]. 

\bibitem{Planck:2018}
Planck Collaboration, Y. Akrami, et al.,
``Planck 2018 results. VI. Cosmological parameters,''
Astron. Astrophys. {\bf 641} (2020) A6, arXiv:astro-ph/1807.06209.

\bibitem{BOSS:2014hwf}
BOSS Collaboration, Timoth\'ee Delubac, et al.,
``Baryon acoustic oscillations in the Ly$\alpha$ forest of BOSS quasars,''
Astron.Astrophys. {\bf 574} (2015) A59, arXiv:astro-ph.CO/1404.1801.


\bibitem{Jaime:2022}
L.G. Jaime, G. Arciniega,
``A unified geometric description of the universe: From inflation to late-time acceleration without an inflaton nor a cosmological constant,''
Phys. Lett. B {\bf 827} (2022) 136939.

\end{thebibliography}
\end{document}